\documentclass[reprint,journal=apchd5,manuscript=article]{achemso}
\usepackage{graphicx}
\usepackage[utf8]{inputenc}

\author{M.\ Bia\l{}ek}
\email{marcin.bialek@fuw.edu.pl}
\affiliation{Institute of Physics, \'Ecole Polytechnique F\'ed\'erale de Lausanne (EPFL), 1015 Lausanne, Switzerland}
\altaffiliation{Centera Laboratories, Institute of High Pressure Physics, Polish Academy of Sciences, Warsaw 01-142, Poland}
\author{W.\ Knap}
\altaffiliation{Centera Laboratories, Institute of High Pressure Physics, Polish Academy of Sciences, Warsaw 01-142, Poland}
\author{J.-Ph.\ Ansermet}
\alsoaffiliation{Institute of Physics, \'Ecole Polytechnique F\'ed\'erale de Lausanne (EPFL), 1015 Lausanne, Switzerland}

\title{Cavity-mediated coupling of antiferromagnetic spin waves}
\date{\today}

\begin{document}
\begin{abstract} 
Coupling of space-separated oscillators is interesting for quantum and communication technologies.  
In this work, it is shown that two antiferromagnetic oscillators placed inside an electromagnetic cavity couple cooperatively to its terahertz modes and, in effect, hybridized magnon-polariton modes are formed.
This is supported by a systematic study of reflection spectra from two parallel-plane slabs of hematite ($\alpha$-Fe$_2$O$_3$), measured as a function of their temperatures and separation distance, and modeled theoretically. The mediating cavity was formed by the crystals themselves and the experiment was performed in a practical distance range of a few millimetres and above room temperature.
Cavity-mediated coupling allows for engineering of complex resonators controlled by their geometry and by sharing properties of their components.
\end{abstract}
\maketitle

\section{Introduction}
Strong coupling between light and matter states gives rise to hybridization into polariton states and Rabi oscillations \cite{Torma14}. Cooperative interaction of light with $N$ resonators increases the splitting by a factor of $\sqrt{N}$, as was observed in a wide range of systems \cite{Raizen89, Khitrova06, Colombe07, Basov16, Bayer17}. With magnetic materials, strong coupling of microwave cavity modes and ferromagnetic resonance was observed as early as 1962 \cite{Roberts62}, but this effect only gained broader attention in the 2010s \cite{Schuster10, Abe11, Huebl13, Zhang14, Tabuchi14, Tabuchi15, Zhang15, Zhang16, Li19, Potts20, Lachance-Quirion20, Li20JAP, Bhoi21}, because of the prospect of using magnetic polaritons in quantum devices \cite{Kasprzak06, Awschalom07, Torma14, Dovzhenko18, Kockum19, Roux20, Yuan22}.
Coupling of matter states mediated by an electromagnetic cavity mode was shown in the case of mechanical oscillators \cite{Spethmann16}. Coupling of two ferromagnets mediated by a cavity mode was recently achieved experimentally using superconducting circuits \cite{Xu19, Li22} and nanostripline antennas \cite{Hanchen22} and was also discussed theoretically \cite{Xu19, Harvey-Collard22, Nair22, Yang22}. Furthermore, different magnon modes can also couple in some ferrimagnetic materials \cite{Liensberger19}.
Most of these experiments require cryogenic temperatures and microwave radiation.
Here, we demonstrate, above room temperature and in the terahertz (THz) range, cavity-mediated coupling of antiferromagnetic magnons in crystals separated by a well-controlled gap.
Mediating cavity is a Fabry-Perot type cavity defined in part by these two crystals themselves.
We used hematite ($\alpha$-Fe$_2$O$_3$) that is characterised by very low spin damping \cite{Lebrun20, Fischer20, Wang21}. Its antiferromagnetic resonance (AFMR) frequency rises with temperature (above room temperature) \cite{Bialek22}. We show that our experimental data can be interpreted with a model based on the input-output theory, classical electrodynamics or a microscopic model. We find that, in the case of our setup, the maximal separation between samples with observable magnon-magnon coupling is in the easily-achievable millimeter range, that is up to ten times the thickness of slabs used in the experiment.

Antiferromagnets are of special interest \cite{Jungwirth18, Hoffman15, Li20Nature} for spintronics and magnon-polariton research because frequencies of magnons in antiferromagnets reach the THz range and some phenomena occur in them that are unavailable in ferromagnets \cite{Li20PRL, Reitz20, Ghosh21}. 
Antiferromagnetic polaritons were achieved mostly at low frequencies \cite{Everts20}, in bulk samples \cite{Grishunin18, Shi20} or via an indirect coupling \cite{Li18, Sivarajah19}, whereas direct strong cavity-magnon coupling at high frequencies was shown only very recently \cite{Bialek21}. One of the reasons is the technical difficulty of constructing THz cavities of high-enough quality factor \cite{Jarc22}, as they need to be much smaller than the ones used in the microwave range. Here, we show that reflection from a Fabry-Perot type cavity shows strong coupling of its modes with AFMR. We show versatility of this scheme of observing magnetic polaritons by achieving coupling of AFMR in two spaced parallel-plane slabs of hematite.

\section{Experimental}
The experiment relies on a continuous-wave THz spectrometer based on frequency extenders linked to a vector network analyzers (VNA), characterized by a very high frequency resolution. Thanks to its very high dynamic range, we can detect AFMR even in transmission though absorptive samples \cite{Caspers16, Bialek19, Bialek20, Zhang20}. In this communication, we used an extender emitting linearly polarized and monochromatic radiation spanning from 0.2 to 0.35 THz. The extender probes $S_{11}$ signal coming back to the emission antenna, thus allowing us to measure reflection at $0$~deg incidence angle without any beam splitters.
\begin{figure}
\begin{center}
\includegraphics[width=0.9\linewidth]{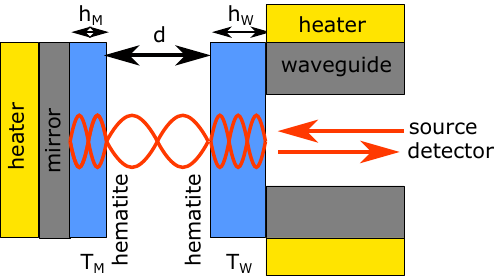}
\caption{{\label{setup}
The measurement system comprises a hematite crystal of thickness $h_M$ placed on a copper mirror and a second hematite crystal of thickness $h_W$ placed at the end of an oversized copper waveguide. We measured reflection in the frequency range of 0.2 to 0.28 THz as a function of gap $d$ between the crystals, imposing a temperature difference $\Delta T=T_M-T_W$, while keeping $T_M+T_W$ constant.}}
\end{center}
\end{figure}

We used two single crystals of natural hematite in [10-10] cut of 0.39,  0.5 mm in thickness and lateral dimensions of 10$\times$10~mm$^2$.
From the extender, the THz beam propagated toward the cavity using an oversized metallic waveguide. At its end, a hematite crystal of $h_W=0.5$~mm thickness was placed (Fig.\ \ref{setup}). This waveguide was mounted between Peltier elements that allowed temperature control of the sample mounted on its end. The second crystal of thickness $h_M=0.39$~mm was placed on a copper mirror and its temperature was controlled with another Peltier element. Temperatures of both crystals were independently controlled and the size $d$ of the air gap between the crystals was controlled with a motorized stage. Thermal transfer due to convection was small for $d>0.2$~mm. In order to reduce possible temperature inhomogeneity of the crystal placed on the end of the waveguide, we chose temperatures as little as possible above room temperature, and the hematite AFMR fell in the spectral range of our extender (i.e. above 0.2 THz).

The detector measured the amplitude and phase of the reflected electric field $E$ at the same polarization angle as that of the emitted beam. Spectra were collected as a function of gap $d$ between the crystals for a fixed temperature difference $\Delta T = T_M-T_W$. Then, both temperatures where changed with a step of 0.25~K, keeping $T_M+T_W$ constant. This step was chosen so that the frequency change was smaller than the line width in the interaction region. Spectra were transformed in the time domain to cut-off reflection coming from the antenna itself and then transformed back into the frequency domain. Finally, magnitude data were divided by a reference spectrum, calculated as a median reflection for all values of $d$ and $\Delta T$. 
The power magnitude data are in dB unit, i.e. they are given as  20log$_{10}|E|$.

\section{Results and analysis}
\begin{figure}
\includegraphics[width=\linewidth]{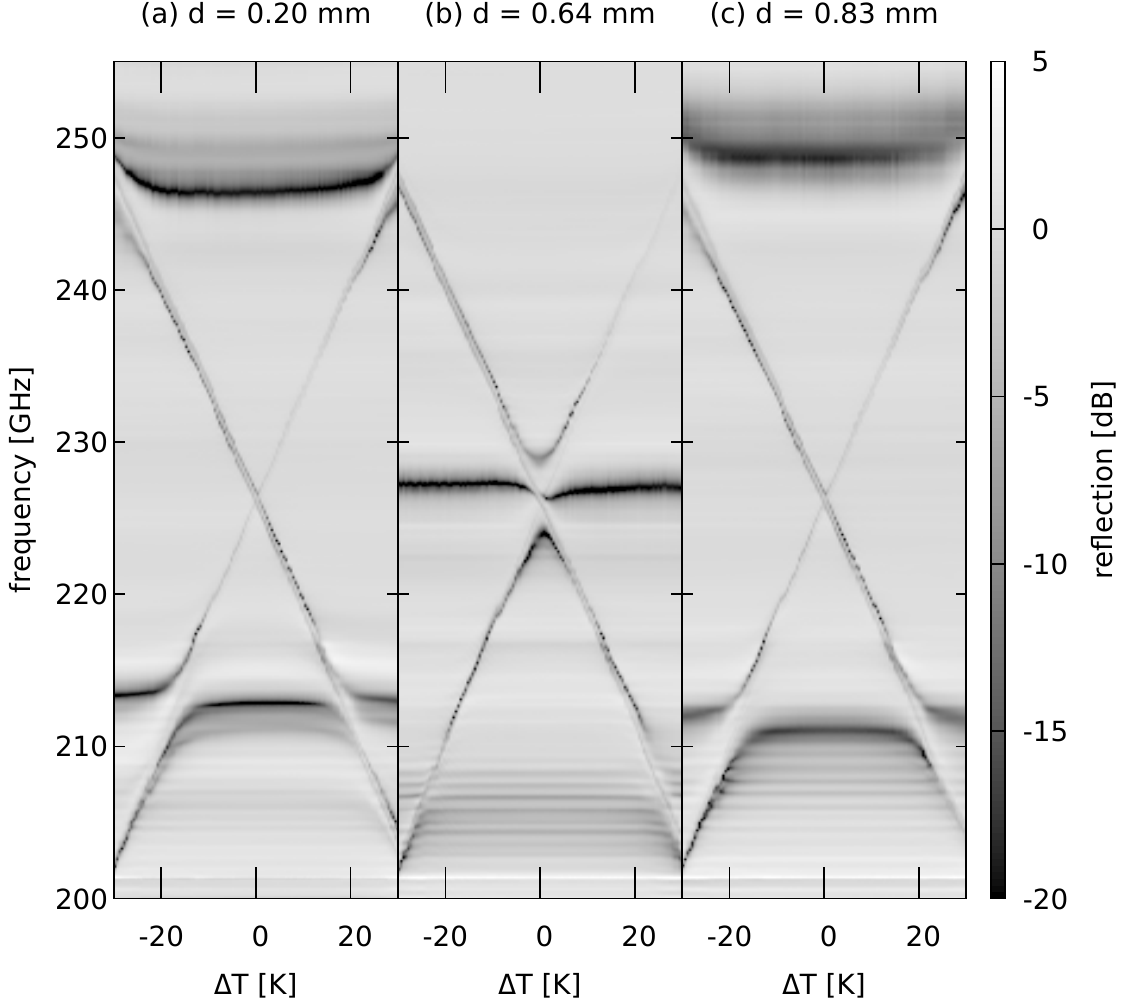}
\caption{\label{mag_T}
 Normalized reflection magnitude at three different lengths of a gap $d$ between the crystals obtained for $(T_M+T_W)/2=336$~K that results in the crossing point at $f\approx227$~GHz.
}
\end{figure}
First, let us discuss our spectra presented as a function of the temperature difference $\Delta T=T_M-T_W$, under the condition of a fixed sum $(T_M+T_W)/2$ (Fig. \ref{mag_T}). Without any interaction between both crystals, one expects a crossing of the AFMR modes in both crystals at $\Delta T=0$. 
For example, such a crossing is observed in Fig. \ref{mag_T}a at about $226$ GHz, a frequency dictated by temperature $(T_M+T_W)/2\approx 336$ K. We also observed two Fabry-Perot cavity modes at about $211$ and $248$ GHz. They are strongly interacting with AFMR in each of the crystals, forming polariton states that are visible as avoided crossings. Results presented in Fig.\ \ref{mag_T}b were obtained for a distance $d$ of $0.64$ mm, chosen because a frequency of a cavity mode (227 GHz) is close to the frequency of the crossing of the AFMR. As discusses below using simulations, the data in Fig.\ \ref{mag_L2}b correspond to strong coherent coupling of the AFMR modes in both crystals. The strength of this coupling is larger than the strengths of the coupling when only one of the crystals is resonant with a cavity mode (Fig.\ \ref{mag_T}a). This  indicates that the resonance in the two crystals are cooperatively coupled to a cavity mode. The right panel (Fig.\ \ref{mag_T}c) shows data obtained for $d=0.83$ mm, which again presents almost unperturbed crossing of the AFMR modes as expected because the cavity mode is not resonant with the AFMR frequency at the crossing. Thus, the coupling between the AFMR in both crystals depends on the distance between the crystals and the strength of this coupling has local maxima when the frequency of one of the cavity modes coincides with the crossing point.

\begin{figure}
\includegraphics[width=\linewidth]{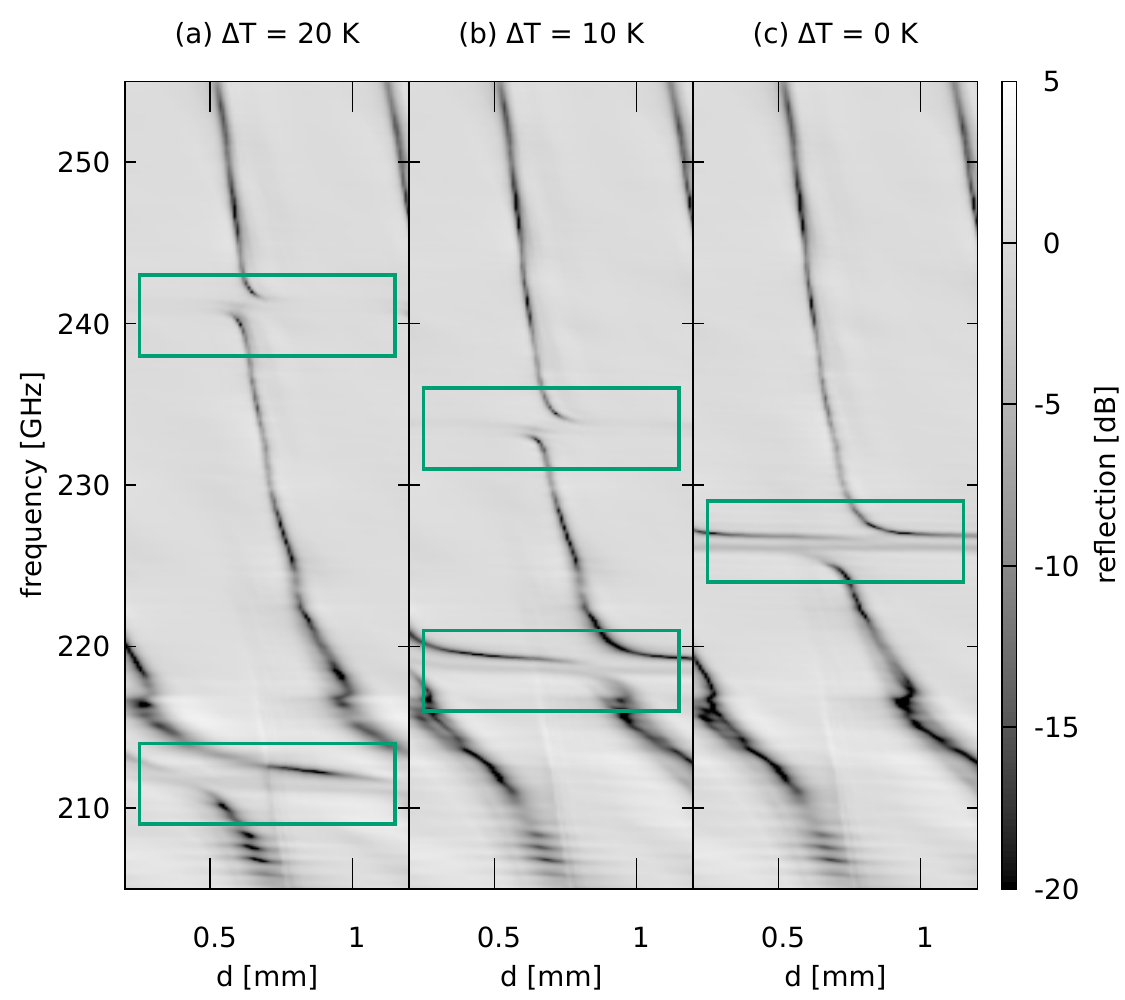}
\caption{\label{mag_d}
Normalized reflection magnitude at three temperature differences $\Delta T$ as a function of gap $d$ between the crystals. Green rectangles mark strong couplings of AFMR with cavity modes. This result comes from the same data set as those in Fig.\ \ref{mag_T}.,
}
\end{figure}
Second, let us examine the data in Fig.\ \ref{mag_d} which present reflection as a function of air gap $d$ for several set temperature differences. Note that the data shown in both Fig. \ref{mag_T} and \ref{mag_d} were extracted from one measurement run during which both $\Delta T$ and $d$ were varied independently. With rising distance, one might expect a monotonic drop of cavity mode frequencies. In Fig.\ \ref{mag_d}, we can see instead formation of avoided crossing and gaps at the  frequencies of AFMR in either one of the crystals (regions bordered with green rectangles). This result is a demonstration of the photonic character of the observed polariton modes, while the result as a function of temperature difference (Fig.\ \ref{mag_T}) shows their magnonic character. In Fig.\ \ref{mag_d} we can see again that the splitting between the polariton modes is larger for $\Delta T=0$ (Fig.\ \ref{mag_d}c) than for couplings when only one of the crystals is on the resonance with the cavity mode (Fig.\ \ref{mag_d}ab).

Let us now discuss our results in the context of three models: a model based on an input-output theory, a model using classical electrodynamics and a microscopic model.

\subsection{Input-output theory}
\begin{figure}
\includegraphics[width=\linewidth]{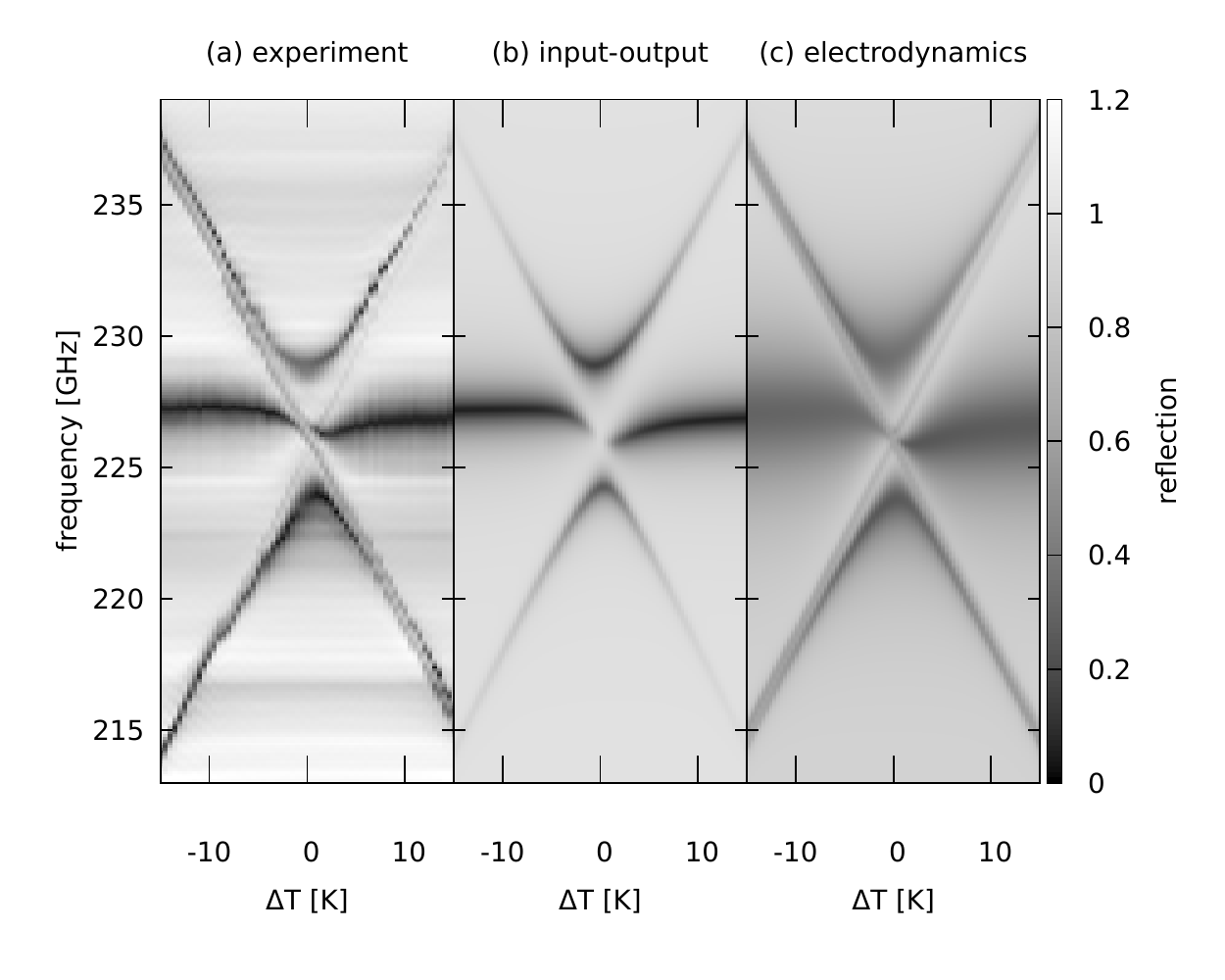}
\caption{\label{IO-ED-fit}(a) Close-up of observed interaction region, (b) fit of an input-output model (Eq.\ \ref{S21-cav}) and (c) predicted reflection using the electrodynamic model.
}
\end{figure}
In the framework of the input-output theory \cite{Schuster10, Harder16, Bialek21}, according to which a strongly coupled system is modelled as an RLC circuit with $L$ containing a resonant term. As shown in Fig.\ \ref{IO-ED-fit}b, we could account for the observed spectra (Fig.\ \ref{IO-ED-fit}a) using:
\begin{equation}
    S_{11} = 1 - a\left(i(f-f_c) - \frac{\kappa}{2}+\sum_{j=M}^{W}\frac{G_j^2}{i(f-f_j(T_j))-\frac{w}{2}}\right)^{-1} 
    \label{S21-cav}
\end{equation}
where, $a=-1.6$~GHz is a parameter describing the coupling of the cavity with the source and the detector, $f_c=225.5$~GHz is the frequency of the cavity mode and $\kappa=3.2$~GHz describes its width, $2G_W=3.6$~GHz and $2G_M=4.8$~GHz are, respectively, splittings between polariton branches for the AFMR in the crystals fixed on the waveguide and on the mirror, and $w=0.7$~GHz describes AFMR width. Normalized coupling strengths are $\eta_W = 2G_W/f_c=0.016$ and $\eta_W = 2G_W/f_c=0.021$. We find that the observed coupling strength between AFMR modes in the two crystals at $\Delta T=0$ is larger than coupling strengths with single crystals and that the magnon-magnon coupling $G_{mm}\approx \sqrt{G_W^2+G_M^2}\approx 6$ GHz. This suggests that spins in both crystals are adding up to interact cooperatively with the cavity field. 


\subsection{Electrodynamics model}
Classical electrodynamics allows analytical calculations of reflection from a system consisting of a series of parallel-plane slabs using characteristic matrix model for isotropic media \cite{BORN}. Although hematite crystal is not isotropic, the off-diagonal terms in $\mu$ or $\epsilon$ are zero around the AFMR frequency \cite{Zavislyak19}. Thus, for our calculations, we can assume an isotropic dielectric constant $\epsilon=18.5$ for hematite ($\epsilon_{air}=1$ for air). 
We take into account the antiferromagnetic resonance in the permeability, writing,
\begin{equation}
    \mu = 1 + \frac{\Delta\mu f_r^2}{f_r^2-f^2-ifw},
\end{equation}
where $f_r$ is its frequency, $w=0.4$ GHz is its width and its strength $\Delta\mu=0.9^{-3}$ as determined in our transmission results \cite{Bialek22}. The strength of the resonance $\Delta \mu$ is responsible for the strength of the coupling of spin waves with electromagnetic waves. In the electrodynamics model, we neglect reflection from the metal mirror, since our calculations show that strong coupling originate from the constructive interference in the crystals themselves. For details please refer to the supplemental material. 

The predicted reflection spectra are shown in Fig.\ \ref{IO-ED-fit}c for gap $d=700$ $\mu$m and crystals thicknesses of $h_M=390$ $\mu$m and $h_W=500$ $\mu$m. The electrodynamics model systematically predicts larger line widths of cavity modes than those observed in the experiment (Fig.\ \ref{IO-ED-fit}a). This could be explained as an effect of elements of the experimental setup that cannot be taken into account by our one-dimensional model or the mirror that we neglect.
In comparison with the model based on the input-output theory, the benefit of using the electrodynamics model is that it naturally takes into account the presence of other cavity modes and can account for absorption by the matter modes. Thus, in the electrodynamics model, we can account for the observed narrow lines in the middle of the interaction region that do not undergo avoided crossing. 

\subsection{Microscopic model}
In order to analyze our data in terms of the quantum mechanism of light-matter coupling \cite{Huebl13}, we estimate the strength of the magnon-photon coupling with
\begin{equation}
G_j = \frac{g_s\mu_B}{2h}\sqrt{\frac{\mu_0h}{2}\frac{V_s}{V_c}\rho f_c},
\label{coupling}
\end{equation}
where, $g_s=2$, $\rho=5\rho_{Fe}$ is the density of spins in hematite, where $\rho_{Fe}=3.987\times10^{28}$~m$^{-3}$ is the density of iron atoms \cite{Pailhe08} and the factor $5$ is the magnetic moment of Fe$^{3+}$ ions \cite{Shull51}. The factor $V_s/V_c$ describes the ratio of the $j$-th crystal volume to the volume of the entire cavity, which in the case of this experiment is of the order of $h_j/(h_1+h_2+d) < 0.5$. The expected splittings calculated this way ($2G_j$) are about an order of magnitude larger than the observed splittings (6 GHz). This discrepancy could be understood with Eq.\ \ref{coupling} as being caused by only a small number of antiferromagnetic magnons interacting with THz photons. Classically, this can be understood as a small spatial overlap of a cavity mode magnetic field with the antiferromagnetic spin wave mode. In our case, determining this overlap is beyond the microscopic model.

\begin{figure}
\includegraphics[width=.7\linewidth]{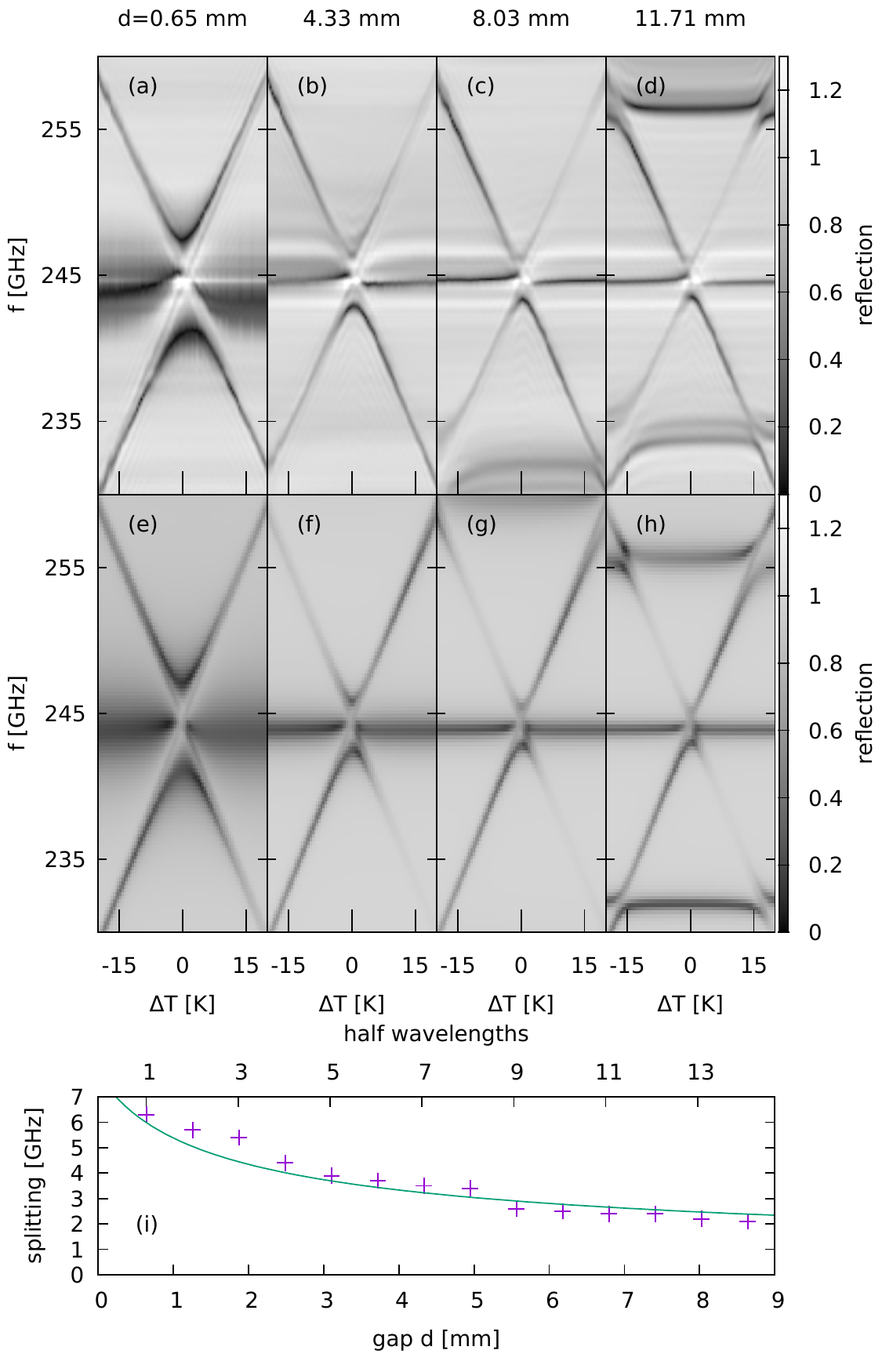}
\caption{\label{mag_L2}
 Top panels (a)-(d) show measured interactions at four selected distances between the crystals, which were equal to integer numbers of half wavelengths. The middle panels (e)-(h) show simulated reflection using the electrodynamics model. The bottom panel (i) shows splitting between the upper and lower magnon-polariton modes as a function of gap $d$ between the crystals; the solid green line is a fit using Eq.\ \ref{Gmm_d}.
}
\end{figure}
In Fig.\ \ref{mag_L2}(a-d), we show results obtained over a large distance $d$ with a step equal to an integer number of half a wavelength of the crossing frequency, which was set to about 244.5 GHz for this experiment. We observed that the coupling drops with increasing gap (Fig.\ \ref{mag_L2}(a-d)) and at above about $d\approx10$ mm it is difficult to recognize the splitting between the three modes (Fig.\ \ref{mag_L2}(d)).
We used the electrodynamics model (Fig.\ \ref{mag_L2}(e-h)) to explain the observed decrease in the splitting with increasing gap (Fig.\ \ref{mag_L2}(i)). This model also  predicts that the quality factor of the cavity mode 
increases with increasing gap. 
Figure \ref{mag_L2}(i) shows the splitting between the low and high modes as a function of gap distance $d$. Assuming that the coupling should be proportional to the square root of the mean density of oscillators in the cavity, we could account for the data using,  
\begin{equation}
    G_{mm}(d)=G_{mm}^{(d=0)}\sqrt{\frac{d_0}{d_0+d}},
    \label{Gmm_d}
\end{equation}
which yielded $d_0 = 0.9\pm0.2$ mm and $G_{mm}(d=0) = (7.9\pm0.6)$ GHz. The fit value $d_0$ coincides with the sum of the actual crystal thicknesses $h_M+h_W=0.89$ mm. This proves the general rule of proportionality of the coupling strength to the square root of the mean density of oscillators in the entire cavity. 


\section{Summary}
We presented systematic experimental and theoretical studies of interaction of two magnetic polariton systems as a function of temperature and distance. Cooperative interaction of the magnetic polariton modes in the space-separated hematite ($\alpha$-Fe$_2$O$_3$) crystals mediated by cavity mode was demonstrated and the characteristic interaction length was determined to be about ten times the sum of their thicknesses. The experiments  were performed above room temperature and distances were in the  mm range, i.e. quite a convenient range. This work shows that cavity-mediated coupling enables controlling antiferromagnetic polaritons by modulating a resonator surroundings, and allows construction of hybrid THz resonators that would share properties of their components.

\begin{acknowledgement}
Support by the Sino-Swiss Science and Technology Cooperation (SSSTC) grant no.\ EG-CN\_02\_032019 is gratefully acknowledged. The VNA and frequency extenders were funded by EPFL and the SNF R'Equip under Grant No.\ 206021\_144983. Partial funding from the European Union’s Horizon 2020 research and innovation programe under the Marie Skłodowska-Curie grant agreement No.\ 847639 and from the Ministry of Education and Science of Poland is acknowledged. This work was also partially supported by the “International Research Agendas” program of the Foundation for Polish Science, co-financed by the European Union under the European Regional Development Fund (No.\ MAB/2018/9).
\end{acknowledgement}

\bibliography{refs}

\end{document}